\documentclass[traditabstract]{aa} 

\usepackage{amsmath}
\usepackage{amssymb}

\usepackage{graphicx}
\usepackage{txfonts}
\usepackage{natbib}
\bibpunct{(}{)}{;}{a}{}{,}

\newcommand{\ergscm}{{\rm erg\,s^{-1} cm}}
\newcommand{\kB}{k_{\rm B}}
\newcommand{\kpc}{\rm kpc}
\newcommand{\yrs}{\rm yrs}
\newcommand{\Gyrs}{\rm Gyrs}
\newcommand{\Geff}{G_{\rm eff}}
\newcommand{\Phieff}{\Phi_{\rm eff}}
\newcommand{\Reff}{R_{\rm e}}
\newcommand{\dotM}{\dot M_{\rm B}^{\rm{e}}}
\newcommand{\dotm}{\dot m}
\newcommand{\vtilde}{\tilde{\upsilon}}
\newcommand{\rhotilde}{\tilde{\rho}}
\newcommand{\csinfq}{c_{\rm s,\infty}^{2}}
\newcommand{\csinf}{c_{\rm s,\infty}}
\newcommand{\xmin}{x_{\rm min}}
\newcommand{\Lambdac}{\Lambda_{\rm c,eff}}
\newcommand{\lambdaceff}{\lambda_{\rm c,eff}}
\newcommand{\lambdac}{\lambda_{\rm c}}

\newcommand{\DMbon}{\dot M_{\rm B}}
\newcommand{\rbon}{R_{\rm B}}
\newcommand{\mue}{\mu}
\newcommand{\mpr}{m_{\rm p}}
\newcommand{\sigmat}{\sigma_{\rm T}}
\newcommand{\mach}{\mathcal{M}}
\newcommand{\csound}{c_{\rm s}}
\newcommand{\rhog}{\rho_{\rm g}}
\newcommand{\rgas}{r_{\rm g}}
\newcommand{\rhogasmean}{\overline{\rho}_{\rm g}}
\newcommand{\NH}{\langle N_{\rm H} \rangle}
\newcommand{\Tvir}{T_{\rm vir}}
\newcommand{\EC}{\dot E_{\rm C}}
\newcommand{\DMstarp}{\dot M_{\ast}^+}
\newcommand{\DMwind}{\dot M_{\ast}^{\rm w}}

\newcommand{\W}{W_{\ast}}
\newcommand{\delt}{\delta t}

\newcommand{\Mto}{M_{\rm TO}(t)}
\newcommand{\DMto}{\dot M_{\rm TO}(t)}
\newcommand{\RSN}{R_{\rm SN}}
\newcommand{\LSN}{L_{\rm SN}}

\newcommand{\NBsun}{\Upsilon_{\rm B \,\odot}}

\newcommand{\rdel}{r_{\Delta}}
\newcommand{\rvir}{r_{\rm vir}}

\newcommand{\rhoh}{\rho_{\rm h}}
\newcommand{\Mh}{M_{\rm h}}
\newcommand{\rh}{r_{\rm h}}
\newcommand{\sigmavirq}{\sigma^2_{\rm vir}}
\newcommand{\vcirc}{\upsilon^2_{\rm c}}

\newcommand{\Mstar}{M_{\ast}}
\newcommand{\Mbh}{M_{\rm BH}}
\newcommand{\Msun}{M_{\odot}}
\newcommand{\Mgal}{M_{\rm inf}}
\newcommand{\Mgas}{M_{\rm g}}
\newcommand{\Minf}{M_{\rm inf}}
\newcommand{\TC}{T_{\rm C}}

\newcommand{\DMgas}{\dot M_{\rm g}}
\newcommand{\DMinf}{\dot M_{\rm inf}}
\newcommand{\DMstar}{\dot M_{\ast}}
\newcommand{\DMbh}{\dot M_{\rm BH}}
\newcommand{\DMesc}{\dot M_{\rm esc}}
\newcommand{\DMbhac}{\dot M_{\rm BH,acc}}
\newcommand{\DMedd}{\dot M_{\rm Edd}}

\newcommand{\Ledd}{L_{\rm Edd}}
\newcommand{\fedd}{f_{\rm Edd}}
\newcommand{\bhstar}{\beta_{\rm BH,\ast}}
\newcommand{\etaesc}{\eta_{\rm esc}}

\newcommand{\tinf}{\tau_{\rm inf}}
\newcommand{\tcool}{\tau_{\rm cool}}
\newcommand{\tdyn}{\tau_{\rm dyn}}
\newcommand{\tesc}{\tau_{\rm esc}}
\newcommand{\tstar}{\tau_{\ast}}

\newcommand{\alstar}{\alpha_{\ast}}

\newcommand{\dr}{{\rm d}r}

\newcommand{\Mg}{M_{\rm g}}
\newcommand{\rg}{r_{\rm g}}
\newcommand{\Rtilde}{\tilde{R}}

\newcommand{\Sigmastar}{\Sigma_{\rm g}(R)}
\newcommand{\MPg}{M_{\rm Pg}(R)}

\newcommand{\LT}{\Lambda (T)}
\newcommand{\tinfall}{t_{\rm inf}}    
\newcommand{\tstarform}{t_{\ast}}               
\newcommand{\tbh}{t_{\rm BH}}                     
\newcommand{\tgas}{t_{\rm g}}                    
\newcommand{\tE}{t_{\rm E}}                      
\newcommand{\B}{ B}

\newcommand{\rev}[1]{{ #1}}
\newcommand{\revle}[1]{{ #1}}

\begin{document}
   \title{One-zone models for spheroidal galaxies\\ with a central supermassive black-hole}

   \subtitle{Self-regulated Bondi accretion}

   \author{E. Lusso$^{1,2}$\thanks{elisabeta.lusso2@unibo.it}
          \and
          L. Ciotti$^{1}$
          }

   \institute{
$^{1}$Dipartimento di Astronomia, Universit\`{a} di Bologna, via Ranzani 1, I-40127 Bologna, Italy.\\
$^{2}$INAF-Osservatorio Astronomico di Bologna, via Ranzani 1, I-40127 Bologna, Italy.\\
             }
   \authorrunning{E. Lusso and L. Ciotti}
   \titlerunning{One-zone models for spheroidal galaxies}
   \date{Accepted, September 27, 2010}

  \abstract
   {By means of a one-zone evolutionary model, we study the co-evolution of supermassive black holes and their host galaxies, as a function of the accretion radiative efficiency, dark matter content, and cosmological infall of gas.
In particular, the radiation feedback is computed by using the self-regulated Bondi accretion.
The models are characterized by strong oscillations when the galaxy is in the AGN state with a high accretion luminosity.
We found that these \rev{one-zone} models are able to reproduce two important phases of galaxy evolution, namely an obscured-cold phase when the bulk of star formation and black hole accretion occur, and the following quiescent hot phase in which accretion remains highly sub-Eddington.
A Compton-thick phase is also found in almost all models, associated with the cold phase.
An exploration of the parameter space reveals that the closest agreement with the present-day Magorrian relation is obtained, independently of the dark matter halo mass, for galaxies with a low-mass seed black hole, and the accretion radiative efficiency $\simeq0.1$.
}


   \maketitle
%

\section{Introduction}

Elliptical galaxies invariably contain central supermassive black holes (SMBHs), and there exists a tight relationship between the characteristic stellar velocity dispersion $\sigma$ (or stellar mass $\Mstar$) of the host system and the SMBH mass $\Mbh$ (e.g., \citealt{mtr+98,fm00,tgb+02,yt02}). These relations clearly indicate a co-evolution of the SMBHs and their host spheroids. Several investigations have been dedicated to this subject, either by using hydrodynamical simulations (e.g., \citealt{co97,co01,co07}; \citealt{2009ApJ...699...89C,2010ApJ...717..708C}) or one-zone models (e.g., \citealt[ hereafter SOCS]{sos05}; \citealt{2008A&A...478..335B}, \citealt{2008arXiv0804.1492M}). In some work, the effect of galaxy merging has also been taken into account (e.g., \citealt{H2005,H2006}).
The main advantage of hydrodynamical models is that complex physical phenomena effects (such as shock waves, jets, radiation transport, etc) can be taken into account. However, the computational time of the simulations force us to search for faster methods that allow a more systematic exploration of the parameter space, which is prohibitive with hydrodynamical simulations.
In this framework, hydro-simulations can be used to set the acceptable range for parameters to be adopted in toy models (e.g., the duty cycle value, see Sect. \ref{The unchanged physics}), while simpler models, such as that used here, are useful to \revle{identify} the most interesting cases that can be simulated in detail with hydrodynamical codes.
\par
\rev{The general idea behind one-zone models is to work with ``average" equations that capture some aspect of a more complicated situation. In practice, some of the equations are exact (such as for example the mass and energy balance equation). On the other hand, some of the physical variables are volume or mass averaged (e.g., the mean gas and temperature of the interstellar medium, respectively), or finally computed at some fiducial radius of the assumed gas distribution. From this point of view, the specific galaxy and dark matter halo profiles do not enter directly into the code, but are needed to obtain realistic mean values to be used in the equations.}
\par
A preliminary investigation of a physically based one-zone toy model has been done in SOCS, where it is shown that for a typical quasar spectral energy distribution (SED) the final $\Mbh$ produced by feedback clearly reproduces the observed $\Mbh$-$\sigma$ relation.
The new models discussed here contain important improvements with respect to SOCS. First of all,  the modelization of accretion onto SMBH. In this paper we examine how the effects of radiation pressure due to the Thomson electron scattering modify in a self-consistent way the spherical Bondi flow.
Another improvement is the treatment of Type Ia supernova rates and mass losses due to the evolution of the galactic stellar population, as we now adopt the \citet{kr05} initial mass function (IMF) coupled with the evolutionary prescription of \citet{mar05}. Finally, the dark matter halo is now described by a finite-mass \citet{1983MNRAS.202..995J} distribution instead of a singular isothermal sphere.
\par
We focus on a scenario in which the masses of the central SMBH and the host galaxy grow in a dark matter halo, which is replenished by accretion of gas of cosmological origin. We follow star formation and we also consider the mass return from the evolving stellar populations. The combined effect of SNIa heating and radiative feedback, during episodes when the luminosity from the central SMBH approaches its Eddington limit, heats and drives much of the remaining gas out of the galaxy, limiting both future growth of the SMBH and future star formation to low levels.
We do not consider the merging phenomenon, and we restrict ourselves to the evolution of an isolated galaxy: it is well known that significant AGN activity may also be present in isolated systems, because of the large amounts of gas \revle{produced by} passively evolving stellar populations (e.g., \citealt{1983ApJ...272..390M,1991ApJ...376..380C,2009ApJ...699..923B,2010arXiv1008.1583K,2010arXiv1009.3265C}).
\rev{More specifically, the one-zone models discussed in this paper are developed as an alternative approach to studying the ``feedback modulated accretion flows'' of \citet{2009ApJ...699...89C,2010ApJ...717..708C}.}
\rev{There are two major differences between the present approach and hydro-simulations. The first, and most obvious, is the impossibility to resolve physical phenomena associated with specific length and timescales. The second is that we consider heating feedback only, while in the current version of the hydrodynamical code by \citet{2009ApJ...699...89C,2010ApJ...717..708C}, both mechanical feedback, and radiation pressure are also considered. 
However, despite these shortcomings, with the present models we can attempt to simulate the process of galaxy formation, which is beyond the current possibilities of the hydro-simulations and at the same time we can explore the parameter space.}
\par
The evolutionary scenario that we consider here addresses several key observational findings. First, that giant ellipticals are old -- they end
their period of vigorous star formation early in cosmic time, since the radiative output from the central SMBHs limits (in cooperation with the
energetic input due to star formation) the gas content to be at levels for which ongoing star formation is minimal. Secondly, gas-rich,
actively star-forming galaxies at redshift $z\sim 3$, including Lyman break galaxies and bright submillimeter SCUBA galaxies, generally
exhibit AGN activity \citep{s+02,a+03,a+09,l+04,2010ApJ...719.1393D}, indicating that their central SMBHs continue to grow. 
This suggests that the formation of a spheroid probably closely preceeds a quasar shining phase, as verified by spectroscopic observations indicating that quasars occupy metal-enriched environments (e.g., \citealt{hf99}, \citealt{2009ApJ...690.1672S}, \citealt{2010arXiv1002.3156W}). 
The redshift evolution of the quasar emissivity and the star formation history of spheroids is thus expected to have evolved roughly in parallel since $z\sim 3$, which is also consistent with observations (e.g., \citealt{hco04}, \citealt{hkb+04}). Among the most important observational predictions of the model is the length of the so-called ``obscured accretion phase" (e.g., \citealt{2004ASSL..308..245C}). This phase is defined as the period of time when a high column density is associated with a high accretion rate onto the central SMBH. We also study the relation between the duration of the obscured phase and the corresponding ``cold phase" (defined by a low mass-weighted gas temperature), and how they depend on the adopted parameters.
Finally, in a large set of models we explore how the final SMBH mass is related to the final stellar mass, as a function of the dark matter halo mass, the amount of cosmological infall and the accretion rate of the gas, the SMBH accretion radiative efficiency, and finally the initial SMBH mass.
We find that the models are in close agreement with the observed Magorrian relation when we assume high efficiencies ($\epsilon\sim 0.1$) and relatively low initial SMBH masses ($\Mbh \lesssim 10^6\Msun$).

This paper is organized as follows. In Sect. 2, we present the physics adopted in the simulations, with special emphasis on the differences from and improvements on SOCS. Section 3 is devoted to the description of the adopted self-regulated accretion model, while in Sect. 4 we present our findings. The main results are summarized and discussed in Sect.~\ref{appendix}, while several useful interpolating functions that we use to compute the stellar mass losses and the SNe Ia rate are presented in the Appendix.

\section{The model}
\label{galmod}
The galaxy model and the input physics adopted for the simulations have been improved with respect to SOCS, in particular the treatment of SNIa heating, the mass return rate from evolving stars, and the accretion onto the central SMBH.
Several aspects of this model were described in SOCS and Ciotti \& Ostriker (2007), and we present here only the relevant modifications.

\subsection{The unchanged physics}
\label{The unchanged physics}

For completeness, we briefly summarize the aspects of the input physics that remain unchanged with respect to SOCS. The differential equation for the gas mass balance is
\begin{equation}
\label{gas}
 \DMgas=\DMinf-\DMstar-\fedd \;\DMbh-\DMesc.
\end{equation}
The first source term on the right-hand side (r.h.s.) describes the cosmological infall in a pre-existing (and time-independent) dark matter halo
\begin{equation}
\label{infall}
 \DMinf=\frac{\Mgal}{\tinf} e^{-t/\tinf},
\end{equation}
where $\Mgal$ is the total gas mass accreted during the simulation (but in general not equal to the final stellar galaxy mass $\Mstar$), and $\tinf$ is the characteristic infall timescale. A more appropriate description of the cosmological gas infall could be obtained by multiplying the r.h.s. of Eq.~(\ref{infall}) by a factor $t/\tinf$ (e.g., \citealt{2009ApJ...697L..38J}), so that the total infalling mass is unaffected, while the infall starts in a more gentle way. For completeness, we performed some simulation with this different description, and we found (as expected) that for reasonable values of $\tinf$ the results are not much affected. Therefore, for consistency with SOCS we retain Eq.~(\ref{infall}). The second source term is the net star-formation rate
\begin{equation}
\label{starform}
 \DMstar=\DMstarp-\DMwind,
\end{equation}
where
\begin{equation}
\label{starformp}
 \DMstarp=\frac{\alstar \Mgas}{\tstar}
\end{equation}
is the instantaneous star-formation rate, and $\DMwind$ is the mass return by the evolving stellar population (Appendix \ref{appendixA1}); following SOCS, in the simulations $\alpha_\ast$ is fixed to be 0.3. The characteristic star-formation timescale $\tstar$ is defined as 
\begin{equation}
\label{tstar}
 \tstar=\max(\tdyn,\tcool),
\end{equation}
where the dynamical timescale $\tdyn$ is given by Eq.~(\ref{tdyn}), and the mean gas cooling time $\tcool$ is estimated to be
\begin{equation}
\label{tcooldef}
 \tcool=\frac{E}{\EC}.
\end{equation}
In the equation above, $\EC$ is the fiducial cooling rate given by
\begin{equation}
\label{collrate}
 \EC=n_{\rm e}n_{\rm p} \LT=\left(\frac{\rhogasmean}{\mpr}\right)^2 \frac{X(1+X)}{2}\LT,
\end{equation}
where $\rhogasmean$ is the instantaneous mean gas density (see Sect. \ref{The Galaxy Model}),
\begin{equation}
\LT={\frac{2.18\times 10^{-18}}{T^{0.6826}}}+2.706\times 10^{-47}T^{2.976}
\;\ergscm
\end{equation}
is the cooling function (\citealt{mb78}; see also \citealt{co01}), and $X=0.7$ is the hydrogen mass abundance (for simplicity we assume complete ionization). In agreement with Eqs.~(\ref{tcooldef}) and (\ref{collrate}), the mean gas internal energy is
\begin{equation}
\label{meaninternalenergy}
 E=  \frac{3 \rhogasmean \kB T}{2 \mu \mpr},
\end{equation}
where $\mu=(0.25+1.5X+0.25Y)^{-1}\simeq 0.62$ is the mean atomic weight.
It follows that
\begin{equation}
\label{tcool}
 \tcool=\frac{8 \pi \rg^3 \kB \mpr }{\mu X(X+1)\Mgas } \frac{T}{\LT},
\end{equation}
where $\Mgas$ is the instantaneous value of the total gas mass and $\rg$ is the gas distribution scale radius (Sect. \ref{The Galaxy Model}).
The third source term is the total accretion rate onto the SMBH
\begin{equation}
\label{DMbh}
 \DMbh=\DMbhac+\bhstar \DMstarp.
\end{equation}
The first term describes gaseous accretion (Sect. \ref{bondimodified}), whereas the second term represents the contribution by the coalescence of stellar remnants of massive stars, as discussed in SOCS.
The numerical coefficient $\fedd\approx 10^{-2}-10^{-3}$ needs to be implemented in the one-zone code (which by definition is unable to model the different spatial scales of the problem) to represent the observed time variation of quasars. In practice, $\fedd$ represents the ``duty-cycle'', and its value is constrained by both observations (e.g. \citealt{yt02}, \citealt{hco04}) and simulations (\citealt{co07}, \citealt{2009ApJ...699...89C,2010ApJ...717..708C}).
When the thermal energy of the interstellar medium (ISM) of the galaxy is high enough, the gas is able to escape from the dark matter potential well, at a fiducial escape rate computed as
\begin{equation}
\label{DMesc}
\DMesc=\left\{
     \begin{array}{l}
      \displaystyle \frac{\Mgas}{\tesc}  \; \;\; \;\; \;\; \; T \geqslant \etaesc\Tvir, \\ \\
      \displaystyle              0   \; \;\;\; \;\; \;\; \;\; \;\; \; T< \etaesc\Tvir.
     \end{array}
\right.
\end{equation}
The parameter $\etaesc$ is of the order of unity, while the expression for $\Tvir$ is given in the next section.
Finally, the escape characteristic timescale is
\begin{equation}
 \tesc=\frac{2\rg}{\csound},
\end{equation}
where $\csound$ is the speed of sound and $\rg$ is the scale length of the gas distribution (Sect. \ref{The Galaxy Model}).
Energy input into the ISM is provided by the thermalization of supernova ejecta (both SNII and SNIa). The treatment of SNII is the same as in SOCS, and the new numerical treatment of SNIa is described in Appendix \ref{appendixA2}. Additional contributions to the ISM energetics come from the thermalization of red giant winds and radiative feedback due to accretion onto the SMBH. As in SOCS, we adopt the average quasar SED obtained from the CRB supplemented by information from individual objects. We recall that the UV and high energy radiation from a typical quasar can photoionize and heat a low density gas up to an equilibrium Compton temperature ($\TC\approx 2\times 10^7$\,K) that exceeds the virial temperatures of giant ellipticals.
Following SOCS, we also consider adiabatic cooling in the case of gas escaping and heating/cooling due to inflow/outflowing galactic gas. The gas temperature is therefore determined at each time-step by integrating the equation of the internal energy per unit volume
\begin{equation}
\label{ergvolume}
 \dot E = \dot E_{\rm H,SN} + \dot E_{\rm H,w} + \dot E_{\rm H,AGN} -\EC +\dot E_{\rm ad}
 +3\frac{\DMinf\lambda \upsilon^2_{\rm esc}-\DMesc\csound^2}{16 \pi \rg^3},
\end{equation}
where $\dot E_{\rm H,SN}$ is the energy due to SNIa and SNII, $\dot E_{\rm H,w}$ describes the thermalization of red giant winds, $\dot E_{\rm H,AGN}$ is the AGN heating, $\dot E_{\rm ad}$ is the adiabatic cooling in the case of galactic winds, and $\lambda$ is a dimensionless parameter ($0.25\leqslant\lambda\leqslant1$). 
Finally, we force the gas to remain above $10^4$ K.

\subsection{The dark matter halo}
\label{The Dark Matter Halo}

For simplicity, the dark matter (DM) halo in our model is the only contributor to the gravity of the galaxy.
In addition, the gas and the stellar density distributions in the simulations are assumed to be proportional to the local (unevolving) DM density distribution, modeled as a Jaffe (1983) profile (Sect. \ref{The Galaxy Model})
\begin{equation}
\label{DMdensity}
 \rhoh(r) = \frac{\Mh}{4 \pi} \frac{\rh}{r^2(\rh+r)^2}.
\end{equation}
For a DM halo total mass $\Mh$, the scale length $\rh$ is fixed following \citet{lanz2004}.
In that paper, 13 massive DM halos obtained from high-resolution cosmological N-body simulations were carefully analyzed.
In particular, for each halo the ``overdensity radius" $\rdel$ (sometimes called virial radius) was determined, and it was shown that $\rdel$ does not differ significantly from the true virial radius $\rvir$ of the system (with a scatter $\lesssim$ 20\%).
Averaging over the 13 clusters reported in Table 2 in \citet{lanz2004}, we found that
\begin{equation}
\label{scalrel}
 \Mh \cong 0.03 \rdel^3,
\end{equation}
where $\Mh$ is in $10^9\Msun$ and $\rdel$ is in $\kpc$ units.
For the Jaffe profile $\rvir=2\rh$, and from Eq.~(\ref{scalrel}) we can link the scale-length $\rh$ to $\Mh$ by assuming $\rvir=\rdel$.
We finally recall that the virial (3D) velocity dispersion associated with Eq.~(\ref{DMdensity}), is given by
\begin{equation}
\label{vcirc}
 \sigmavirq=\frac{G \Mh}{2\rh}.
\end{equation}
The halo mean circular velocity is estimated as $\vcirc= 2\sigmavirq$, and the dynamical time required in Eq.~(\ref{tstar}) is therefore given by
\begin{equation}
\label{tdyn}
 \tdyn=\frac{2\pi\rh}{\upsilon_{\rm c}}.
\end{equation}

\subsection{The galaxy model}
\label{The Galaxy Model}

One of the new aspects of the model evolution investigated in this paper is the time extent of the Compton thin and Compton-thick phases, as defined at the end of this section.
In SOCS, the adopted gas density distribution was a singular isothermal sphere and the major drawback of this choice for the evaluation of average quantities is the divergence of the total mass.
For this reason, we now use the more realistic Jaffe density profile, i.e., we assume that at each time the gas density distribution is
\begin{equation}
\label{rhogas}
 \rhog(r) = \frac{\Mgas}{4 \pi} \frac{\rgas}{r^2(\rgas+r)^2},
\end{equation}
where $\rgas$ is the gas scale radius and $\Mgas$ is the instantaneous value of the total gas mass, obtained by integrating Eq. (\ref{gas}).
For simplicity, $\rg=\rh$ during the whole simulation. However, it is possible to generalize the present approach and consider a time-dependent $\rg$.

The mean gas density, which is needed in Eqs.~(\ref{collrate}) and (\ref{meaninternalenergy}), is evaluated to be the instantaneous mean value within the half-mass radius $\rg$
\begin{equation}
\label{gasmeandensity}
\rhogasmean = \frac{3 \Mgas}{8\pi \rg^3},
\end{equation}
while the virial gas temperature $\Tvir$ is obtained from the hydrostatic equilibrium and the Jeans equations solved in the gravitational potential of the DM halo neglecting the self-gravity of the gas. Simple algebra then shows that
\begin{equation}
\label{virialT}
 \Tvir = \frac{\mu \mpr G \Mh}{6\kB \rg}.
\end{equation}

To quantify the relative importance of the optically thin and thick phases, the code computes the gas column density as
\begin{equation}
\label{NHdef}
 \NH = \frac{\MPg}{2\pi R^2},
\end{equation}
where $\MPg$ is the projected gas mass enclosed in a circle of radius $R$, and the factor of 2 at the denominator takes into account that only one side of the gas column actually obscures the center.
Therefore, the fiducial column density depends not only on the total gas mass, but also on the aperture radius adopted.
To simulate observational work (e.g. \citealt{1999ApJ...522..157R}), we decided to define $R$ to be one hundredth of the effective radius $\Reff$\footnote{For the Jaffe model, $\Reff\cong 0.74\rg$.}, i.e., in Eq. (\ref{NHdef}) $R\cong0.0074 \rg$ .
From Eq.~(\ref{Mprojsol})
\begin{equation}
 \NH \cong 32.7 \frac{\Mgas}{\rg^2},
\end{equation}
and $\rg$ is fixed during the simulation, $\NH$ depends only on $\Mgas$.
We refer to the ``thick" phase if $\NH \geq 10^{24}$ cm$^{-2}$, while galaxies with column density $\NH \geq 10^{22}$ cm$^{-2}$ are considered obscured.

\subsection{The code}
\label{The Code}

The simulations are performed by numerical integration of the previous evolutionary differential equations using a forward Euler scheme. The numerical code is based on the code developed in SOCS. The time step is defined as the minimum among the characteristic times associated with the different physical processes
\begin{equation}
\label{stepdef}
 \delta t=\B \min \left(\tinfall,\tstarform,\tgas,\tbh,\tE \right),
\end{equation}
where the different subscripts indicate the specific aspect of the physics involved and, in general, for a quantity $X_i$, $t_i\equiv X_i/|\dot X_i|$. The dimensionless coefficient $\B\leq 1$ is used to improve the code accuracy. 
We performed several test simulations and we found that $\B \cong0.1$ leads to rapid convergence and excellent agreement with results obtained in SOCS, when used with their model galaxies and input physics.


\section{Self-regulated Bondi accretion}
\label{bondimodified}

\begin{figure}
 \resizebox{\hsize}{!}{\includegraphics{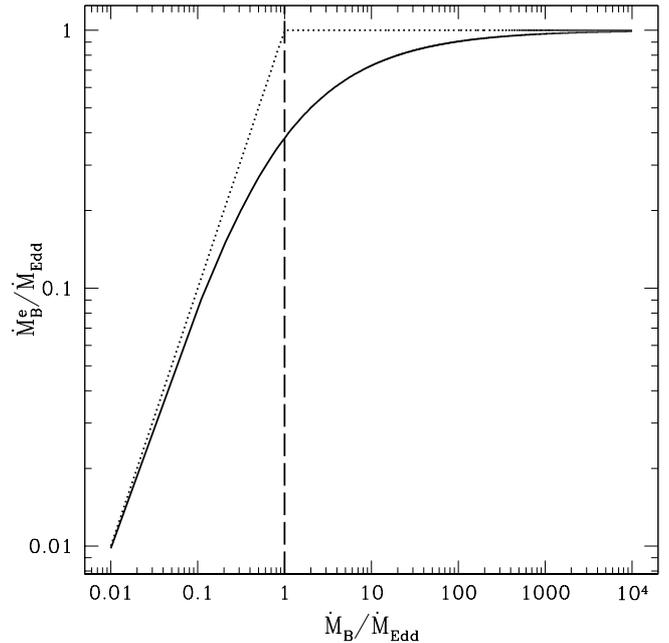}}
 \caption{The self-regulated Bondi accretion rate $\dotM$ obtained from Eq.~(\ref{DMbhacc}) is represented by the heavy solid line. The dotted line is the accretion rate determined by Eq.~(\ref{olddmacc}). Note how for very low and very high classical accretion rates $\DMbon$, the two prescriptions coincide.}
 \label{bondimodif}
\end{figure}

As often assumed in the literature, the SMBH accretion rate is determined as the minimum between the Bondi accretion rate $\DMbon$ and Eddington limit $\DMedd$
\begin{equation}
\label{olddmacc}
 \DMbhac=\min (\DMedd,\DMbon).
\end{equation}
In the previous formula, the Eddington accretion rate is 
\begin{equation}
 \DMedd\equiv\frac{\Ledd}{\epsilon c^2},
\end{equation}
where $0.001\lesssim\epsilon\lesssim0.1$ is the accretion efficiency, and 
\begin{equation}
 \Ledd=\frac{4\pi c G \Mbh \mue \mpr}{\sigmat}
\end{equation}
is the Eddington luminosity and $\sigmat$ is the Thomson cross-section.
The classical Bondi accretion rate is
\begin{equation}
\label{DMbon}
  \DMbon=\lambdac 4 \pi~G^2 \Mbh^2~\rho_{\infty} \csinf^{-3},
\end{equation}
where $\rho_{\infty}$ and $\csinf$ are the gas density and the speed of sound at infinity, respectively, and $\lambdac$ is a dimensionless coefficient of the order of unity (e.g., \citealt{1952MNRAS.112..195B}, \citealt{1999agnc.book.....K}). In eq~(\ref{DMbon}), radiative effects are not taken into account, so that in Eq.~(\ref{olddmacc}) there is a sharp transition between the pure hydrodynamical and radiation-dominated regimes. Fortunately, in the optical thin regime dominated by electron scattering, it is possible to extend the classical Bondi accretion solution and take into account the radiation pressure effects, so that the transition between Bondi-limited and Eddington-limited accretion can be described in a more consistent way. After solving this problem, we discovered that is had already been fully described by \citet{taamfu91} and \citet{fukue01}, and here we just summarize the main features of the modified accretion model. The result is obtained basically, by noting that in the optically thin regime the radiation pressure scales as $1/r^2$, thus reducing the effective gravitational force at each radius by the same amount. By imposing self-consistentcy, i.e. by requiring that the effective accretion rate $\dotM$ is determined by the effective gravity, one finally obtains the following expression for the modified coefficient
\begin{equation}
\lambdaceff=\lambdac \left(1-\frac{\dotM}{\DMedd}\right)^2.
\end{equation}
Therefore, the self-consistent Bondi accretion rate satisfies the equation
\begin{equation}
\label{DMbhaccpre}
 \dotM= \DMbon \left(1-\frac{\dotM}{\DMedd}\right)^2,
\end{equation}
that can be solved for $\dotM$. After discarding the unphysical solution, we have 
\begin{equation}
\label{DMbhacc}
 \dotm\equiv\frac{\dotM}{\DMedd}=\frac{1}{2}\left[2+r-\sqrt{4r+r^2} \right],\quad r\equiv\frac{\DMedd}{\DMbon}.
\end{equation}
For $r \rightarrow 0$ (high accretion rates) $\dotm \rightarrow 1$, so that $\dotM$ tends to the Eddington accretion, while for $r \rightarrow \infty$ (low accretion rates) $\dotm \rightarrow 0$. The full solution is indicated in Fig.~\ref{bondimodif} by the heavy solid line: we note how Eq.~(\ref{olddmacc}) overestimates accretion onto SMBH in the range $0.1 \leqslant\DMbon/\DMedd\leqslant 100$.

\begin{figure}
\centering
 \resizebox{\hsize}{!}{\includegraphics[width=1.6cm]{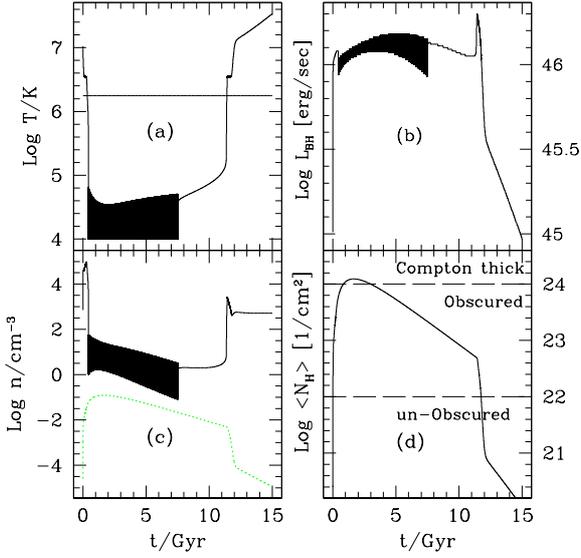}}
 \caption{The time evolution of relevant quantities of the RM. \textit{Panel a}: gas temperature; the virial temperature of the galaxy is the horizontal dashed line. \textit{Panel b}: bolometric accretion luminosity. \textit{Panel c}: gas density at the Bondi radius (black line) and the mean gas density (green dotted line). \textit{Panel d}: mean gas column density $\NH$ at aperture radius of $0.088$ kpc.}
 \label{modelRM}
\end{figure}

\begin{figure}
 \resizebox{\hsize}{!}{\includegraphics[width=1.5cm]{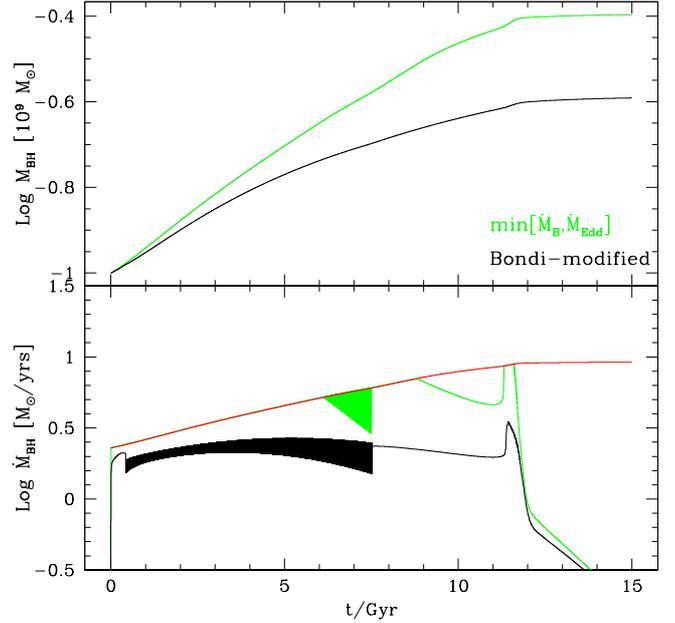}}
 \caption{The SMBH accretion history of the RM. \textit{Top panel}: time evolution of the SMBH mass as computed according to SOCS (green line) and with the Bondi-modified accretion (black line). \textit{Bottom panel}: SMBH accretion rate from SOCS (green line) and from our work (black line). The red solid line is the Eddington accretion rate for the SOCS model.}
 \label{mbhRM}
\end{figure}

\section{Results}
\label{results}

In the following sections, we present the main properties of a set of simulations, focusing on three important issues and comparing the new results with those in SOCS. 
However, before discussing the whole set of the new models, in the next section we present in detail the evolution of a representative model and three of its possible variants.

\begin{table*}[ct]
\caption[]{Final properties of the RM and its variants at 15 Gyrs.}
\label{table:4models}
\begin{tabular}{llllllllccclc}
\hline\hline
\noalign{\smallskip}
Model & $\Mstar$ & $M_{\rm{esc}}$ & $\Mbh$ & $\Mbh/\Mstar$ & $\log \NH$& $\Delta t_{\rm cold}^1$& $\Delta t_{\rm cold}^2$&$\Delta t_{\rm CT}$ & $\Delta t_{\rm obs}$& $\Delta t_{\rm unobs}$ \\
\noalign{\smallskip}
\hline
\noalign{\smallskip}
 RM       &   93.22 & 6.56  & 0.256 &   $2.75 \times 10^{-3}$  & 20.05 & 10.42 &  7.94 & 2.17  &  9.57    &   3.26   \\
 RM$_1$   &   92.47 & 6.62  & 0.945 &   $1.02 \times 10^{-2}$  & 20.04 & 10.37 &  7.91 & 2.13  &  9.56    &   3.31   \\
 RM$_2$   &   42.65 & 7.21  & 0.214 &   $5.01 \times 10^{-3}$  & 19.63 &  8.42 &  5.78 & 0.00  & 10.33    &   4.67   \\
 RM$_3$   &  192.97 & 6.69  & 0.321 &   $1.67 \times 10^{-3}$  & 20.46 & 12.04 &  9.69 & 4.92  &  8.17    &   1.91   \\
 RM$_4$   &   90.59 & 5.55  & 0.288 &   $3.18 \times 10^{-3}$  & 22.81 & 14.20 & 10.47 & 0.00  & 14.99    &   0.01   \\
\noalign{\smallskip}
\hline\hline\noalign{\smallskip}
\end{tabular}

\begin{list}{}{Note -- All masses are in $10^9\Msun$ units. $\Mbh$ is the final black hole mass; $\Mstar$ is the final stellar mass;  $M_{\rm{esc}}$ is the total escaped gas mass. The average column density is in cm$^{-2}$. $\Delta t_{\rm CT}$, $\Delta t_{\rm obs}$, and $\Delta t_{\rm unobs}$ represent the durations (in Gyr) of the Compton-thick phase, of the obscured phase, and the unobscured phase, respectively. Finally, $\Delta t_{\rm cold}^1$ and $\Delta t_{\rm cold}^2$ are the durations (in Gyr) measured assuming threshold temperatures of $10^5$ K and $5\times 10^4$ K, respectively.}
\item
\end{list}
\end{table*}

\subsection{The evolution of the reference model and some of its variants}
\label{The evolution of the reference model}

Figure \ref{modelRM} shows the time evolution of important quantities of our galaxy reference model (RM). This model is characterized by a DM halo of total mass $\Mh=4\times 10^{11}\,\Msun$, which corresponds to a halo scale length $\rh=11.86$ kpc and $\Reff=8.83$ kpc, a fiducial circular velocity calculated according to Eq.~(\ref{vcirc}) of $\sim 270$ km/sec, and a characteristic infall time 2 Gyr. The assumed total mass of the cosmological gas infall is $\Mgal=10^{11}\,\Msun$, corresponding to a dark-to-total mass ratio of 80\% (provided that all the infalling gas forms stars). Other simulation parameters are $\alpha_*=0.3$, $\beta_{\rm BH,*}=1.5\times10^{-4}$, $\epsilon=0.1$, $\eta_{\rm SN}=0.85$, $\eta_{\rm esc}=2$, and a stellar mass-to-light ratio of 5 (see Eq.~[\ref{RSN}]). As in SOCS, the initial SMBH mass is $10^8\,\Msun$; the duty circle, $\fedd$, is fixed at 0.005.

As can be seen from the comparison of Fig.~\ref{modelRM} with Figs.~7 and 8 in SOCS, the global evolution of the new models is qualitatively very similar to those in SOCS.
In particular, the time evolution of the model, from the beginning up to $\sim9$ Gyrs, is characterized by a cold phase (we assume that a cold phase corresponds to $T_{\rm gas}\leq 10^5$K) of high density and low temperature. The gas density at the Bondi radius remains between 1 and $10^2$ particles per cm$^3$, while the mean gas density is $\sim10^{-1}-10^{-2}$ particles per cm$^3$. At the beginning of the cold phase, about 2.17 Gyrs are spent in the Compton-thick phase (see Fig.~\ref{modelRM}d).
The remaining part of the cold phase is obscured, while the accretion luminosity remains high, at $L_{\rm BH}\gtrsim10^{46}$ erg/sec.
As soon as the mean gas density decreases (Fig.~\ref{modelRM}c green dotted line) the cooling becomes inefficient, the gas heating dominates, and the temperature increases to the virial temperature. The total duration of the cold phase is $\sim 9$ Gyr, which corresponds to a decrease in $\NH$ of about 2 orders of magnitude. The obscured phase lasts $\sim 9.6$ Gyr and the average column density is then around $\NH \sim10^{20}$ cm$^{-2}$.
The durations of the cold and the obscured phases are very similar, and indeed, the two phases are related. Until the gas density is high, the gas can radiate efficiently the energy input due to the AGN, and its temperature remains below $10^5$ K. Because of this large amount of cold gas, the star formation proceeds at high rates, until most of the gas is consumed. At this point, the radiative cooling of the gas becomes inefficient and the heating due to the AGN causes the gas temperature to increase. The combined effects of the gas consumption and the increase temperature stop the star formation.
\rev{A specific feature of the cold phase can be seen in Fig.~\ref{modelRM}, where characteristic temperature oscillations are apparent. These oscillations, already presented and discussed in SOCS, are due to the combined effect of gas cooling and AGN feedback. They terminate when the gas density falls below some threshold determined by the cooling function and because less and less gas is produced by stars and accreted by the galaxy, while SNIa heating declines less strongly. When the density is high (at early times), the cooling time is instead, very short, and AGN heating is radiated efficiently. The two competitive effects produce the temperature (and density) oscillations. In the hydro-simulations, the spatial and temporal structures of these oscillations is quite complicated, as feedback and cooling act on several different spatial and temporal scales (from a month to 10 Myr). In the one-zone models, the oscillations are instead dependent on the ``duty-cycle'' parameter $\fedd$ in Eq.~(\ref{gas}), which is constrained by observational and theoretical studies (see Sect. \ref{The unchanged physics}).}

To summarize, the observational properties of the RM would correspond to a system initially Compton-thick, that then switches to an obscured phase, and in the past 3.2 Gyrs the galaxy is unobscured.
The SMBH accretion history is shown in Fig.~\ref{mbhRM}, where it is compared to that predicted by the SOCS formulation in a model otherwise identical to the RM. The major difference between the two models is in the final value of $\Mbh$: in particular, Eq.~(\ref{olddmacc}) would predict a present-day $\Mbh$ a factor $\sim 1.5$ higher than that for modified accretion.
This is because the accretion in SOCS remains for almost all of the galaxy life, at the Eddington limit, while in the new treatment the self-regulation maintains the accretion at lower rates.

Before presenting the results of a global exploration of the parameter space, we focus on a few obvious questions. For example, what happens if we keep all the model parameters fixed and reduce the radiative accretion efficiency? What happens if we increase or decrease the total mass of infalling gas? Or if we double the gas infalling time?
Of course, these simple examples do not cover all the possible cases. However, these models will provide a guide as we investigate the parameter space. The relevant properties of the additional ``reference" models are listed in Table \ref{table:4models}.

\begin{figure}
 \resizebox{9.5cm}{!}{\includegraphics{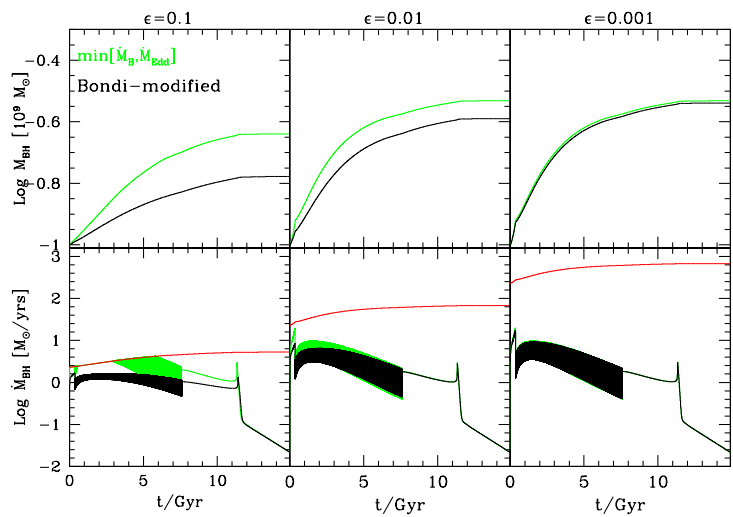}}
\caption{Evolution of the SMBH mass and accretion rate computed according to Eq.~(\ref{olddmacc}) (green line, SOCS) and Eq.~(\ref{DMbhaccpre}) (black line) for a model with $\Minf=1.25\times10^{10}\Msun$, $\Mh=5\times10^{10}\Msun$, and initial SMBH mass of $10^8\Msun$. From left to right: $\epsilon=0.1$, $\epsilon=0.01$, and $\epsilon=0.001$. The red line represents the Eddington limit for the SOCS model.}
\label{mbh3x3}
\end{figure}

The first variant of the RM model is model RM$_1$ obtained by reducing the radiative accretion efficiency from $\epsilon=0.1$ to $\epsilon=0.001$.
Overall, this reduced efficiency model passes through the same evolutionary phases as model RM: an initial Compton-thick phase, followed by an obscured phase, and finally a low-density unobscured phase.
The initial cold high density phase is also very similar to that of RM in Fig.~\ref{modelRM}.
The most important and expected difference is in the final mass of SMBH, which is higher by a factor of $\approx 3.5$ than the RM. A reduction in the radiative efficiency also produces a slightly larger amount of escaped gas, because of the shorter cold phase.
As a consequence, a larger amount of gas produced by the evolving stars is lost from the galaxy during the low-density, hot phase. 
The effect of a reduction in $\Minf$ is explored in model RM$_2$, while all the others parameters are the same as in RM.
Qualitatively, the evolution is again very similar to that of RM in Fig.~\ref{modelRM}.
The main difference is the absence of the initial Compton-thick phase because of the lower gas density.
Both the final stellar mass and final SMBH mass are lower than in model RM, as can be seen from Table~\ref{table:4models}.
Overall, model RM$_2$ does not display remarkable or unexpected \revle{properties}. The only noticeable aspect is that the escaped mass is higher than in model RM, but this is again due to the shorter cold phase. 


A complementary model to RM$_2$ is RM$_3$, where we double the value of $\Minf$ while maintaining all the other parameters identical to those of model RM. The qualitative evolution is again similar to that of RM in Fig.~\ref{modelRM}, but, at variance with RM$_2$, the Compton-thick phase is now present.
Not surprisingly, the Compton-thick phase in model RM$_3$ extends for a longer period ($\sim 5$ Gyrs) than in the other variants of RM (see Table~\ref{table:4models}), and its total mass of new stars is also the highest.
However, the total mass that escapes is not as high as one would expect, the infall mass being a factor of 2 higher than in the RM yet the escaped mass being almost the same. The main reason for this behaviour is the very massive star formation, which is almost double that of RM.
\rev{We note that model RM$_3$ is the closest, in the RM family, to the galaxy models studied in the hydrodynamical simulations of \citet{2009ApJ...699...89C,2010ApJ...717..708C}, as far as the final stellar mass and the final $\Mbh$ are concerned.}

We finally discuss model RM$_4$, which is identical to model RM, but has twice as long an infall time.
The main effects are the longest cold phase in RM family (see Table~\ref{table:4models}), and the absence of the initial Compton-thick phase. This latter characteristic is due to the time dilution of the infalling gas density, which prevents the possibility of reaching very high column density values.
Overall, the final SMBH mass is not affected significantly by the extended infall phase, but its escaped gas masses is quite high because the total mass in new stars is (as expected) lower than in the RM model.

To summarize, these preliminary experiments have revealed that, at fixed dark matter halo, sensible variations in the input parameters do not produce remarkable different results.
The relevant differences are found mainly in the amount of star formation and the different durations of the cold and obscured phases, while the final SMBH mass appears to be mainly affected by the value of the radiative efficiency $\epsilon$.
\begin{table*}[ct]
\caption[]{Final properties of models with self-regulated Bondi accretion with an initial SMBH mass of $10^8\Msun$. All masses are in units of $10^9\Msun$. }
\label{table:models}
\begin{tabular}{lll|llllllcclc}
\hline\hline
\noalign{\smallskip}
$\Mgal$ & $\alpha$& $\epsilon$ &  $\Mstar$ & $M_{\rm{esc}}$ & $\Mbh$ &$\Mbh/\Mstar$ & $\log \NH$&  $\Delta t_{\rm cold}^1$&  $\Delta t_{\rm cold}^2$&$\Delta t_{\rm CT}$ & $\Delta t_{\rm obs}$& $\Delta t_{\rm unobs}$ \\
\noalign{\smallskip}  
\hline\hline          
\noalign{\smallskip}
\multicolumn{12}{c}{$\rm M_1=10$} \\
\hline
\noalign{\smallskip}
  2.5  & 0.25  & 0.1  &  2.40 & 0.07 & 0.132 &$5.51\times 10^{-2}$ & 18.95 & 10.74 & 8.39 & 0.00  & 11.29   & 3.71   \\
  2.5  & 0.25  & 0.01 &  2.38 & 0.07 & 0.154 &$6.45\times 10^{-2}$ & 18.94 & 10.69 & 8.35 & 0.00  & 11.23   & 3.77   \\
  2.5  & 0.25  & 0.001&  2.37 & 0.07 & 0.158 &$6.67\times 10^{-2}$ & 18.94 & 10.68 & 8.35 & 0.00  & 11.22   & 3.78   \\
  5.0  & 0.5   & 0.1  &  4.88 & 0.07 & 0.144 &$2.95\times 10^{-2}$ & 19.36 & 12.18 & 9.96 & 0.00  & 12.70   & 2.30   \\
  5.0  & 0.5   & 0.01 &  4.85 & 0.07 & 0.177 &$3.65\times 10^{-2}$ & 19.35 & 12.13 & 9.92 & 0.00  & 12.64   & 2.36   \\
  5.0  & 0.5   & 0.001&  4.84 & 0.07 & 0.184 &$3.80\times 10^{-2}$ & 19.35 & 12.12 & 9.92 & 0.00  & 12.64   & 2.36   \\     
\noalign{\smallskip}  
\hline\hline                
\noalign{\smallskip}
\multicolumn{12}{c}{$\rm M_2=50$} \\
 \hline
\noalign{\smallskip}                                                                                                                                            
  12.5  & 0.25  & 0.1  &  11.23 & 1.20 & 0.167 &$1.49\times 10^{-2}$ & 19.43 & 10.51 & 8.11 & 0.00 & 11.45  &  3.55  \\ 
  12.5  & 0.25  & 0.01 &  11.12 & 1.21 & 0.257 &$2.32\times 10^{-2}$ & 19.42 & 10.45 & 8.07 & 0.00 & 11.39  &  3.61  \\ 
  12.5  & 0.25  & 0.001&  11.07 & 1.23 & 0.289 &$2.61\times 10^{-2}$ & 19.42 & 10.45 & 8.06 & 0.00 & 11.38  &  3.62  \\ 
  25.0  & 0.5   & 0.1  &  24.10 & 0.79 & 0.190 &$7.90\times 10^{-3}$ & 19.84 & 12.15 & 9.87 & 2.31 & 10.55  &  2.14  \\ 
  25.0  & 0.5   & 0.01 &  23.96 & 0.80 & 0.328 &$1.37\times 10^{-2}$ & 19.83 & 12.10 & 9.84 & 2.29 & 10.51  &  2.20  \\ 
  25.0  & 0.5   & 0.001&  23.92 & 0.80 & 0.370 &$1.55\times 10^{-2}$ & 19.83 & 12.10 & 9.84 & 2.29 & 10.52  &  2.20  \\ 
\noalign{\smallskip}  
\hline\hline                
\noalign{\smallskip}
\multicolumn{12}{c}{$\rm M_3=100$} \\
\hline
\noalign{\smallskip}
  25.0 & 0.25  & 0.1  & 23.07 & 1.82 & 0.189 &$8.18\times 10^{-3}$ & 19.64 &  10.53  & 8.11  & 0.00  & 11.51  & 3.49   \\
  25.0 & 0.25  & 0.01 & 22.91 & 1.83 & 0.346 &$1.51\times 10^{-2}$ & 19.63 &  10.48  & 8.07  & 0.00  & 11.45  & 3.55   \\
  25.0 & 0.25  & 0.001& 22.84 & 1.84 & 0.414 &$1.81\times 10^{-2}$ & 19.63 &  10.48  & 8.07  & 0.00  & 11.45  & 3.55   \\
  50.0 & 0.5   & 0.1  & 48.34 & 1.51 & 0.220 &$4.55\times 10^{-3}$ & 20.05 &  12.14  & 9.84  & 3.33  &  9.58  & 2.08   \\
  50.0 & 0.5   & 0.01 & 48.10 & 1.52 & 0.452 &$9.40\times 10^{-3}$ & 20.04 &  12.09  & 9.80  & 3.32  &  9.55  & 2.14   \\
  50.0 & 0.5   & 0.001& 48.00 & 1.53 & 0.536 &$1.12\times 10^{-2}$ & 20.04 &  12.09  & 9.81  & 3.31  &  9.55  & 2.14   \\
\noalign{\smallskip}  
\hline\hline                
\noalign{\smallskip}
\multicolumn{12}{c}{$\rm M_4=500$} \\
\hline
\noalign{\smallskip}
125.0  & 0.25  & 0.1  & 115.84 & 8.91 & 0.273 &$2.36\times 10^{-3}$ & 20.11 & 10.37 & 7.87 & 2.51   &  9.33  &  3.15 \\
125.0  & 0.25  & 0.01 & 115.38 & 8.87 & 0.776 &$6.73\times 10^{-3}$ & 20.10 & 10.32 & 7.84 & 2.50   &  9.29  &  3.21 \\
125.0  & 0.25  & 0.001& 114.95 & 8.97 & 1.104 &$9.61\times 10^{-3}$ & 20.10 & 10.32 & 7.84 & 2.48   &  9.31  &  3.21 \\
250.0  & 0.5   & 0.1  & 240.57 & 9.03 & 0.349 &$1.45\times 10^{-3}$ & 20.52 & 12.01 & 9.65 & 5.14   &  8.01  &  1.84 \\
250.0  & 0.5   & 0.01 & 239.84 & 9.06 & 1.048 &$4.37\times 10^{-3}$ & 20.51 & 11.98 & 9.63 & 5.13   &  7.98  &  1.89 \\
250.0  & 0.5   & 0.001& 239.42 & 9.12 & 1.407 &$5.88\times 10^{-3}$ & 20.52 & 11.98 & 9.63 & 5.12   &  7.99  &  1.89 \\
\noalign{\smallskip}  
\hline\hline                
\noalign{\smallskip}
\multicolumn{12}{c}{$\rm M_5=1000$} \\
\hline
\noalign{\smallskip}
250.0 & 0.25  & 0.1  & 222.85 & 26.74 & 0.350 &$1.57\times 10^{-3}$ & 20.31 & 10.09  & 7.53 & 3.31  &  9.18   &  2.50   \\
250.0 & 0.25  & 0.01 & 222.06 & 26.68 & 1.216 &$5.48\times 10^{-3}$ & 20.30 & 10.05  & 7.51 & 3.30  &  9.08   &  2.61   \\
250.0 & 0.25  & 0.001& 220.80 & 27.16 & 1.994 &$9.03\times 10^{-3}$ & 20.30 & 10.04  & 7.50 & 3.28  &  9.10   &  2.62   \\
500.0 & 0.5   & 0.1  & 471.83 & 27.47 & 0.476 &$1.01\times 10^{-3}$ & 20.72 & 11.85  & 9.44 & 5.78  &  7.77   &  1.45   \\
500.0 & 0.5   & 0.01 & 470.52 & 27.66 & 1.613 &$3.43\times 10^{-3}$ & 20.71 & 11.82  & 9.43 & 5.77  &  7.71   &  1.52   \\
500.0 & 0.5   & 0.001& 469.62 & 27.85 & 2.336 &$4.97\times 10^{-3}$ & 20.71 & 11.83  & 9.43 & 5.77  &  7.72   &  1.52   \\
\noalign{\smallskip}
\hline\hline\noalign{\smallskip}
\end{tabular}

\begin{list}{}{Note -- The halo mass $\Mh$ is ($10^{10}, 5\times 10^{10}, 10^{11}, 5\times 10^{11}, 10^{12}$)~$\Msun$ for models $M_1, M_2, M_3, M_4$, and $M_5$, respectively. The total infall gas mass is $\Mgal=\alpha \Mh$. $\Mbh$ is the final SMBH mass (also with the contribution of the initial SMBH mass). $\Mstar$ is the final stellar mass. $M_{\rm{esc}}$ is the total escaped gas mass. The average column density is in $\rm cm^{-2}$. $\Delta t_{\rm CT}$, $\Delta t_{\rm obs}$, and $\Delta t_{\rm unobs}$ represent the duration (in $\Gyrs$) of the Compton-thick phase, the obscured phase, and the unobscured phase, respectively. Finally, $\Delta t_{\rm cold}^1$ and $\Delta t_{\rm cold}^2$ are the durations (in Gyr) measured assuming threshold temperatures of $10^5$ K and $5\times 10^4$ K, respectively.}\item
\end{list}
\end{table*}
\begin{table*}[ct]
\caption[]{Final properties of models with self-regulated Bondi accretion with an initial SMBH mass of $10^5\Msun$. All masses are in units of $10^9\Msun$.}
\label{table:models5}
\begin{tabular}{lll|llllllcclc}
\hline\hline
\noalign{\smallskip}
$\Mgal$ & $\alpha$& $\epsilon$ &  $\Mstar$ & $M_{\rm{esc}}$ & $\Mbh$ &$\Mbh/\Mstar$ & $\log \NH$&  $\Delta t_{\rm cold}^1$&  $\Delta t_{\rm cold}^2$&$\Delta t_{\rm CT}$ & $\Delta t_{\rm obs}$& $\Delta t_{\rm unobs}$ \\
\noalign{\smallskip}  
\hline\hline          
\noalign{\smallskip}
\multicolumn{12}{c}{$\rm M_1=10$} \\
\hline
\noalign{\smallskip}
  2.5  & 0.25  & 0.1  & 2.43 & 0.06 & 0.002 &$6.72\times 10^{-4}$ & 18.98 &  10.94 &   8.50 &0.00 & 11.51  &  3.49 \\
  2.5  & 0.25  & 0.01 & 2.41 & 0.07 & 0.026 &$1.06\times 10^{-2}$ & 18.94 &  10.70 &   8.37 &0.00 & 11.24  &  3.76 \\
  2.5  & 0.25  & 0.001& 2.36 & 0.07 & 0.071 &$3.02\times 10^{-2}$ & 18.93 &  10.64 &   8.31 &0.00 & 11.18  &  3.82 \\
  5.0  & 0.5   & 0.1  & 4.93 & 0.07 & 0.003 &$6.39\times 10^{-4}$ & 19.39 &  12.36 &  10.06 &0.00 & 12.91  &  2.09 \\
  5.0  & 0.5   & 0.01 & 4.88 & 0.07 & 0.044 &$9.01\times 10^{-3}$ & 19.35 &  12.11 &   9.90 &0.00 & 12.62  &  2.38 \\
  5.0  & 0.5   & 0.001& 4.81 & 0.07 & 0.113 &$2.35\times 10^{-2}$ & 19.34 &  12.09 &   9.88 &0.00 & 12.60  &  2.40 \\     
\noalign{\smallskip}  
\hline\hline                
\noalign{\smallskip}
\multicolumn{12}{c}{$\rm M_2=50$} \\ 
\hline
\noalign{\smallskip}                                                                                                                                            
  12.5  & 0.25  & 0.1  & 11.32 & 1.17 & 0.006 &$5.51\times 10^{-4}$ & 19.45  & 10.61 &  8.16 &0.00  & 11.55  & 3.45   \\ 
  12.5  & 0.25  & 0.01 & 11.23 & 1.19 & 0.071 &$6.36\times 10^{-3}$ & 19.43  & 10.47 &  8.09 &0.00  & 11.39  & 3.61   \\ 
  12.5  & 0.25  & 0.001& 11.10 & 1.20 & 0.192 &$1.73\times 10^{-2}$ & 19.42  & 10.44 &  8.06 &0.00  & 11.37  & 3.63   \\ 
  25.0  & 0.5   & 0.1  & 24.19 & 0.78 & 0.014 &$5.60\times 10^{-4}$ & 19.86  & 12.23 &  9.92 &2.32  & 10.63  & 2.05   \\ 
  25.0  & 0.5   & 0.01 & 24.06 & 0.79 & 0.130 &$5.41\times 10^{-3}$ & 19.83  & 12.09 &  9.84 &2.32  & 10.48  & 2.21   \\ 
  25.0  & 0.5   & 0.001& 23.90 & 0.79 & 0.297 &$1.24\times 10^{-2}$ & 19.83  & 12.09 &  9.83 &2.28  & 10.51  & 2.21   \\ 
\noalign{\smallskip}  
\hline\hline                
\noalign{\smallskip}
\multicolumn{12}{c}{$\rm M_3=100$} \\
\hline
\noalign{\smallskip}
  25.0 & 0.25  & 0.1  & 23.18 & 1.79 & 0.012 &$5.31\times 10^{-4}$ & 19.65  & 10.60  & 8.14  & 0.00  & 11.58   & 3.42  \\
  25.0 & 0.25  & 0.01 & 23.06 & 1.81 & 0.119 &$5.16\times 10^{-3}$ & 19.63  & 10.49  & 8.09  & 0.00  & 11.45   & 3.55  \\
  25.0 & 0.25  & 0.001& 22.87 & 1.82 & 0.300 &$1.31\times 10^{-2}$ & 19.63  & 10.47  & 8.07  & 0.00  & 11.44   & 3.56  \\
  50.0 & 0.5   & 0.1  & 48.44 & 1.50 & 0.026 &$5.42\times 10^{-4}$ & 20.06  & 12.19  & 9.86  & 3.34  &  9.64   & 2.02  \\
  50.0 & 0.5   & 0.01 & 48.24 & 1.52 & 0.212 &$4.40\times 10^{-3}$ & 20.04  & 12.08  & 9.80  & 3.33  &  9.52   & 2.15  \\
  50.0 & 0.5   & 0.001& 48.00 & 1.52 & 0.452 &$9.42\times 10^{-3}$ & 20.04  & 12.08  & 9.80  & 3.31  &  9.55   & 2.14  \\
\noalign{\smallskip}  
\hline\hline                
\noalign{\smallskip}
\multicolumn{12}{c}{$\rm M_4=500$} \\
\hline
\noalign{\smallskip}
125.0  & 0.25  & 0.1  & 116.05 & 8.82 & 0.059 &$5.07\times 10^{-4}$ & 20.12  &    10.39  &   7.88   &  2.52   &    9.35   &    3.13  \\
125.0  & 0.25  & 0.01 & 115.67 & 8.85 & 0.407 &$3.52\times 10^{-3}$ & 20.10  &    10.33  &   7.85   &  2.51   &    9.28   &    3.21  \\
125.0  & 0.25  & 0.001& 115.21 & 8.86 & 0.868 &$7.54\times 10^{-3}$ & 20.10  &    10.32  &   7.84   &  2.49   &    9.29   &    3.22  \\
250.0  & 0.5   & 0.1  & 240.70 & 9.02 & 0.123 &$5.10\times 10^{-4}$ & 20.53  &    12.03  &   9.66   &  5.14   &    8.03   &    1.82  \\
250.0  & 0.5   & 0.01 & 240.15 & 9.03 & 0.681 &$2.84\times 10^{-3}$ & 20.51  &    11.97  &   9.63   &  5.14   &    7.97   &    1.89  \\
250.0  & 0.5   & 0.001& 239.57 & 9.06 & 1.238 &$5.17\times 10^{-3}$ & 20.51  &    11.98  &   9.63   &  5.13   &    7.98   &    1.89  \\
\noalign{\smallskip}  
\hline\hline                
\noalign{\smallskip}
\multicolumn{12}{c}{$\rm M_5=1000$} \\
\hline
\noalign{\smallskip}
250.0 & 0.25  & 0.1  & 222.92 & 26.81 & 0.114 &$5.12\times 10^{-4}$ & 20.31  & 10.10   &  7.54 &   3.31  &     9.21  &     2.48  \\
250.0 & 0.25  & 0.01 & 222.41 & 26.71 & 0.741 &$3.33\times 10^{-3}$ & 20.30  & 10.05   &  7.51 &   3.31  &     9.08  &     2.61  \\
250.0 & 0.25  & 0.001& 221.73 & 26.64 & 1.504 &$6.78\times 10^{-3}$ & 20.30  & 10.05   &  7.51 &   3.30  &     9.07  &     2.63  \\
500.0 & 0.5   & 0.1  & 471.96 & 27.48 & 0.238 &$5.05\times 10^{-4}$ & 20.72  & 11.86   &  9.45 &   5.78  &     7.78  &     1.44  \\
500.0 & 0.5   & 0.01 & 471.01 & 27.55 & 1.167 &$2.48\times 10^{-3}$ & 20.71  & 11.82   &  9.43 &   5.77  &     7.71  &     1.52  \\
500.0 & 0.5   & 0.001& 470.12 & 27.60 & 2.021 &$4.30\times 10^{-3}$ & 20.71  & 11.82   &  9.43 &   5.77  &     7.71  &     1.52  \\
\noalign{\smallskip}
\hline\hline\noalign{\smallskip}
\end{tabular}
\newline{Note -- See Note in Table \ref{table:models} for description.}
\end{table*}

\subsection{Exploring the parameter space}
\label{Exploring the parameters space}
We now present the general results (summarized in Tables \ref{table:models} and \ref{table:models5}) of our examination of the parameter space. The SMBH initial mass in the models in Table \ref{table:models} is $10^8\Msun$, while in Table \ref{table:models5} it is $10^5\Msun$. From the astrophysical point of view, the first choice mimics a scenario in which the central SMBHs are already quite massive at the epoch of galaxy formation, while in the second case the main growth occurs with galaxy formation. Each of the 5 main families of models ($M_1$, ..., $M_5$) consists of galaxy models characterized by the same dark halo mass, $\Mh$, ranging from $10^{10}\Msun$ (the $M_1$ family) up to $10^{12}\Msun$ (the $M_5$ family). In each family, we have investigated 6 models with different values of infalling gas-to-dark matter ratio $\alpha=\Mgal/\Mh$, and finally different radiative efficiencies $\epsilon$ in the self-regulated Bondi accretion.
In practice, in each family we explore the effects of different total infalling gas mass and different efficiencies (spanning the commonly accepted range of values) at fixed $\Mh$.

Almost independently of $\Mh$ and the initial value of $\Mbh$, we can recognize some general trends.
For example in each group of 3 models characterized by identical parameters but decreasing $\epsilon$, it is apparent how the total amount of stars formed decreases at decreasing $\epsilon$, while the final $\Mbh$ increases.
In the total mass budget of the galaxy, an important quantity is the total mass of gas ejected, $M_{\rm{esc}}$. As expected, we find that massive galaxies, with larger infall gas masses also eject more mass. However, in all cases, the escaped gas mass is much lower than the final stellar mass, i.e., $\sim10\%$ of it.

Finally, higher final gas masses are obtained, at fixed $\Mh$, for higher $\Minf$ and lower efficiencies, because of the less effective feedback.
We note however that the product of $\epsilon$ and the mass accreted by the SMBH decreases with decreasing $\epsilon$, i.e., even though the accreted mass is higher, the integrated energy output is lower, and so is the feedback effect\footnote{The higher SMBH masses account for the slightly lower final stellar masses. The mass conservation of the code is quite remarkable, considering the amount of input physics involved ($\Mgal=\Mstar+M_{\rm{esc}}+\Mbh$).}.

A qualitative illustration of the effect of the reduction in $\epsilon$ on the SMBH accretion history is shown in Fig.~\ref{mbh3x3}.
When the efficiency decreases, the final mass of the SMBH increases \revle{and, we indicate by the green line, the corresponding evolution of identical models but with the SOCS treatment}, i.e., where the accretion is determined by the minimum of $\DMbon$ and $\DMedd$.
As expected, the differences in $\Mbh$ reduce with decreasing $\epsilon$, because the Eddington accretion regime in the SOCS models (when the major differences from the Bondi-modified case are established), becomes less and less important, and $\dotM$ approches $\DMbon$. 

All models have a transition phase from obscured to unobscured. The Compton-thick phase is present in almost all models that have a cold phase (18 of 30 simulations); however, in the less massive set of models (M$_{1}$) the Compton-thick phase is absent. 
\rev{This is consistent with the scenario in which the majority of the most massive black holes spend a significant amount of time growing in an earlier obscured phase (see \citealt{2010ApJ...719.1315K,2010arXiv1009.2501T}).}

Finally, the mean SMBH-to-star ratio is not far from the value inferred from the present-day \citet{mtr+98} relation, though on the high mass side.
This is not surprising, because of our use of a one-zone model.
In any case, we emphasize that the SMBH feedback in the models was able to remove (in combination with star formation and/or gas escape) most of the infalling gas (which, if accreted onto the central SMBH in a cooling-flow like solution, would lead to the final SMBH masses being $\sim 2$ orders of magnitude higher than the observed ones).

\begin{figure}
 \resizebox{\hsize}{!}{\includegraphics[width=1.5cm]{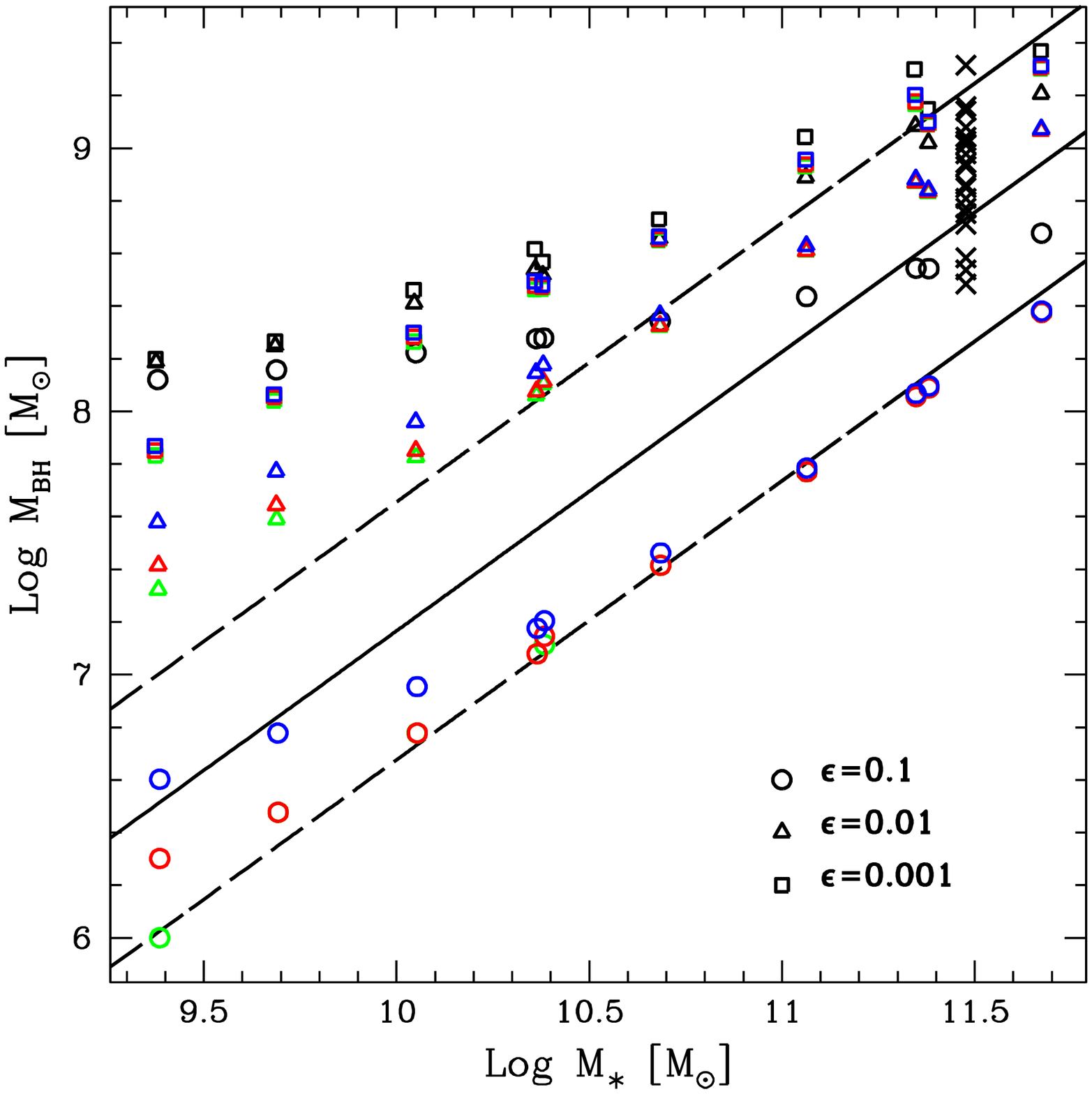}}
 \caption{Final SMBH mass versus final stellar mass for all the explored models. The solid line is the \citet{2003ApJ...589L..21M} best-fit of the Magorrian (1998) relation, and the two dashed lines represent the associated 1-$\sigma$ deviation. Different colors indicate different initial masses of the $\Mbh$, i.e. $10^3\Msun$ (green), $10^5\Msun$ (red, models in Table \ref{table:models5}), $10^6\Msun$ (blue), and $10^8\Msun$ (black, models in Table \ref{table:models}). Different symbols identify the adopted value for radiative efficiency: $\epsilon=0.1$ (circles), $\epsilon=0.01$ (triangles), and $\epsilon=0.001$ (squares).\rev{The black crosses are the hydrodynamical models in Table 1 of Ciotti et al. (2010).}}
 \label{magorrianrelation}
\end{figure}

Independent of the particular characteristics of the single models, Fig.~\ref{magorrianrelation} clearly illustrates that the final SMBH masses, in models with a quite high initial $\Mbh$ (Table \ref{table:models}), are higher than implied by the observed Magorrian relation. For this reason, we explored other families of models, in which the initial mass of the SMBH was reduced to $10^6\Msun$, $10^5\Msun$, and $10^3\Msun$.
From Fig.~\ref{magorrianrelation}, it is apparent how galaxy models with initial SMBH masses $\lesssim10^6\Msun$ (circles) closely agree to within $1\sigma$ of the observed dispersion with observations, but only when the radiative efficiency is high, i.e. $\epsilon=0.1$.
It is even more remarkable that the final SMBH mass is proportional to the final $\Mstar$, as all the models started with the same initial SMBH mass.
This strongly indicates that co-evolution is the most plausible explanation of the proportionality between $\Mstar$ and $\Mbh$, in line with other observational evidence (e.g., see \citealt{hco04} and references therein).
\par
\rev{To compare our present findings with those of a complementary hydrodynamical approach, in Fig.~\ref{magorrianrelation} we plot the results of the high-resolution hydro-simulations developed by Ciotti et al. (2010, Table 1, Cols. 5 and 6) for a representative galaxy with stellar mass of $\sim3\times 10^{11}\Msun$. The final $\Mbh$ are represented by crosses, and the different (luminosity weighted) radiative efficiencies are in the range $0.003\leq\epsilon\leq0.133$. We note that the crosses indicate almost all the hydrodynamical models that have been studied in detail so far, and this shows the importance of one-zone models as a complementary approach to exploring the parameter space. We also note that \revle{the final mass of the SMBHs, as computed in the hydrodynamical simulations, lead to an accurate reprodution of the} Magorian relation but the comparison with the one-zone models is delicate. In the hydrodynamical simulations (at variance with the one-zone models), the initial phases of galaxy formation are not simulated, and the focus is on the maintaining low SMBH masses in the presence of stellar mass losses that, if accreted onto the central SMBH as in an undisturbed cooling flow, would produce a final BH mass of about a factor of $\sim$100 higher than the observed ones. For this reason, in the hydrodynamical models the galaxy is already assumed to have formed, and the initial mass of the SMBH is just slightly lower than the mass predicted by the Magorrian relation.}

\section{Conclusions}

This paper is a natural extension of a previous paper by Sazonov et al. (2005).
From a technical point of view, several aspects of the input physics have been improved.
In particular, we have adopted a different description of the accretion rate, which now follows the self-consistent modified Bondi theory (\citealt{taamfu91}; \citealt{fukue01}), instead of the minimum between the Eddington and the (classical) Bondi rate.
Moreover, the SNIa rate is now computed using a high-precision multi-exponential approximation for the time-kernel in the convolution integral of the star formation rate, instead of the standard power-law. This avoids the need to store the entire star formation history and permits more rapid numerical simulations.
The time-dependent mass return rate from the evolving stellar population is now computed using the \citet{kr05} initial mass function and the mass return rate as a function of the stellar mass given by \citet{mar05}, instead of the Ciotti et al. (1991) formulae.
For the stellar mass-return rate, we also numerically implemented a multi-exponential fit.
Finally, the dark-matter potential well (and the galaxy stellar distribution) are now described as Jaffe (1983) models.
\par
From the astrophysical point of view, in addition to the standard outputs considered in SOCS (e.g. final SMBH mass, final stellar mass, etc), we now also focus on the duration of the ``cold phase", of the ``obscured phase", and the ``Compton thick phase" as fiducially computed using the gas temperature and the column density of cold gas.
The main parameters adopted to fix a model are the dark-matter halo mass (distributed to reproduce cosmological expectations), the amount of gas deemed to flow onto the dark matter halo, and  finally the radiative accretion efficiency. As a separate parameter, we also consider the initial mass of the central SMBH.
After some preliminary model exploration and verifying that the new treatment, when applied to the SOCS galaxy model reproduces the previous results, we performed a significant exploration of the parameter space.
\par
Our main results can be summarized as follows:
\begin{enumerate} 
  \item Almost independently of the dark matter halo mass and the radiative efficiency, the computed galaxy models have an initial phase that extends for some Gyrs, in which the gas temperature is low and the gas density is high. These galaxies would be defined obscured quasars, were we to measure their gas column density within an aperture radius of the order of $\Reff/100$. 

 \item At late times, all the SMBHs are found to be in a low accretion state without much feedback, and the ISM is overall optically thin. The galactic ISM is at about the virial temperature of the dark matter potential well and should be emitting in X-rays, as found for elliptical galaxies in the local universe.

 \item Interestingly, we have found that only a specific class of models, are found to agree with the observed Magorrian relation, i.e. only those with low initial SMBH masses ($\lesssim 10^6\Msun$) and high radiative accretion efficiency, $\epsilon\sim 0.1$. Higher initial SMBH masses, or lower radiative efficiencies lead to final SMBH masses that are too high. Therefore, this result implies that the seed SMBHs should be quite small, but that their mass accretion should occur \revle{mainly with high radiative efficiency}, in agreement with observational findings.

 \item In addition, we have also showed how the self-regulated Bondi accretion recipe can be easily implemented in numerical codes, and how it leads to a lower SMBH mass accretion than the more common "on-off" Eddington regulation.

\end{enumerate}
\par
A series of final comments are in order.
The first is that the presented \rev{one-zone} models seem to be able to reproduce, without much fine tuning, two different phases of galaxy evolution, namely a first obscured phase where much of the SMBH accretion and star formation occurs, followed by a hot phase in which SMBH accretion is highly sub-Eddington and star formation is (almost) entirely absent.
\rev{The robustness of the two-phase evolution, characterized both by one-zone and hydrodynamical simulations is mainly due to the combined effect of (1) the secular decrease in the mass return rate from the evolving stellar populations, (2) the time dependence of the SNIa heating (after the first Gyrs of evolution), and (3) the cooling function, which are identical in the hydrodynamical and the one-zone models. The increase in the specific ISM heating with increasing time, and that a substantial degassing occurs only when the gas temperature is roughly higher by a factor of $\sim4$ than the virial temperature, leads to the appearance of the two phases in the two types of model.}

\rev{The second is that to reproduce the present-day Magorrian relation, the optimal combination of parameters is a quite low initial SMBH mass and a quite high radiative efficiency. We propose that this represents a useful constrain of semi-analytical investigations (also in the context of high-z galaxy merging). As a side-product of this work, we also anticipate that the presented multi-exponential fit for stellar evolution and SNIa heating will be useful in both hydrodynamical and semi-analytical works.}

As a final comment, we point out a major improvement in the code.
In a future study, we will relax the assumption of a fixed gas density distribution, imposing a time dependence of $\rg(t)$ as a function of the gas thermal content at that given time.
We expect that this additional ingredient will cause the models to become more sensitive to the adopted parameters on the one hand by increasing the gas density during the cold phases (hence producing a stronger feedback by decreasing the value of the ionization parameter and increasing the mass accretion rate), and on the other hand by decreasing the gas density during the hot phase, so reducing further the Eddington ratios at late times.  We naturally, also expect that both the star-formation and accretion histories, as well as the durations of the obscured and Compton-thick phases to be affected.
\begin{acknowledgements}
The anonymous referee is acknowledged for his/her comments that greatly improved the paper.
We thank A. Comastri and J.P. Ostriker for useful discussions, and F. Marinacci for suggestions on the numerical code.
The project is supported by ASI-INAF under grants I/023/05/00.
\end{acknowledgements}

\appendix
\label{appendix}

\section{Interpolating functions for stellar evolution}
\begin{table*}[ct]
\begin{center}
\caption{Multi-exponential expansion parameters (in $\Gyrs$) for stellar mass losses and SNIa rate. Note that for SNIa the coefficients $a_i$ are dimensionless.}
\label{table:fit}
\begin{tabular}{llllllllllll}
\hline
\hline
\noalign{\smallskip}
  $a_0$ & $a_1$ & $a_2$ & $a_3$ & $a_4$ & $a_5$ & $b_0$ & $b_1$ & $b_2$ & $b_3$ & $b_4$ & $b_5$ \\
\noalign{\smallskip}  
\hline\hline                
\noalign{\smallskip}               
\multicolumn{11}{c}{Stellar mass return} \\  
\noalign{\smallskip}                                                             
\hline                                                                              
\noalign{\smallskip}                                                                
 16.38 & 1.93 & 0.38 & -- & -- & -- & 1.93 & 0.19 & 0.03 & -- & -- & -- \\
\noalign{\smallskip}  
\hline \hline               
\noalign{\smallskip}               
\multicolumn{11}{c}{SNIa} \\  
\noalign{\smallskip}                                                             
\hline                                                                              
\noalign{\smallskip}                                                                
 0.18 & $3.22\times 10^{-2}$ & $7.15\times 10^{-3}$ & $1.84\times 10^{-3}$ & $5.22\times 10^{-4}$ & $1.34\times 10^{-4}$ & 7.43 & 1.00 & 0.22 & $5.79\times 10^{-2}$ & $1.72\times 10^{-2}$ & $5.28\times 10^{-3}$\\
\hline\hline
\noalign{\smallskip}
\end{tabular}
\end{center}
\end{table*}

\subsection{Stellar mass losses}
\label{appendixA1}

The general expression for the stellar mass-return rate from an evolving stellar population is
\begin{equation}
\label{dMwinddef}
 \DMwind(t)=\int^t_0 \DMstarp(t') \W (t-t') {\rm d}t',
\end{equation}
where $\DMstarp$ is the instantaneous star-formation rate and $\W (t-t')$ is the normalized stellar death rate for a stellar population of age $t-t'$. 
For an initial mass function (IMF) of unitary total mass and age $t$,
\begin{equation}
\label{Wdef}
 \W(t)={\rm IMF}[\Mto]\vert\DMto\vert \Delta M^{\rm w}[\Mto],
\end{equation}
where $\Mto$ is the mass of stars entering the turn-off at time $t$, and $\Delta M^{\rm w}$ their mass loss. Following \citet{co07}, we assume that
\begin{equation}
\Delta M^{\rm w}=\left\{
  \begin{array}{l}
       0.923 M-0.48,\quad 0.08 \leqslant M \leqslant 8.5\\
       M -1.4, \quad\quad\quad\;\;\,8.5 < M \leqslant 40\\
       M/2, \quad\quad\quad\quad\quad M > 40.
  \end{array}
\right.
\end{equation}
In our models, we adopt a Kroupa (2001) IMF, with a minimum mass of 0.08 $\Msun$ and a maximum mass of 100 $\Msun$, while $\DMto$ is taken from Maraston (2005)
\begin{equation}
\label{Mto}
     \begin{array}{l}
 \log \Mto\cong2.982+0.213 \log t -0.108 (\log t)^2 + \\ \quad\quad\quad\quad\quad\;\,0.006 (\log t)^3.
      \end{array}
\end{equation}
In the formula above, $t$ is in yrs and $M_{\rm TO}$ in solar masses.
The main computational problem posed by the evaluation of the integral in Eq.~(\ref{dMwinddef}) is that, in principle, the entire history of $\DMstarp(t')$ must be stored, which requires a prohibitively large amount of memory and computational time.
In the special case of an exponential time dependence of $\W$, this problem can be fortunately avoided, as shown in \citet{co01,co07}.
In the present case, $\W$ is not an exponential function, being more similar to a power law.
However, we assume that a fit in terms of exponentials has been obtained, i.e.
\begin{equation}
\label{W}
 \W(t)= \sum_{i} \frac{e^{-\frac{t}{b_i}}}{a_i},
\end{equation}
where the timescales $a_i$ and $b_i$ are known.
From the above equation and Eq.~(\ref{dMwinddef}), it follows that
\begin{equation}
\label{dMwinddef_coeff}
 \DMwind(t)=\sum_i \frac{F_i(t)}{a_i}
\end{equation}
where
\begin{equation}
\label{F}
 F_i(t) \equiv \int_0^t \DMstarp(t') e^{-\frac{t-t'}{b_i}} {\rm d}t'.
\end{equation}
We divide the integral into
\begin{equation}
\label{dMwind_0}
 F_i(t)=\int^t_{t-\delt} \DMstarp(t') e^{-\frac{t-t'}{b_i}} {\rm d}t'+\int^{t-\delt}_0 \DMstarp(t') e^{-\frac{t-t'}{b_i}} {\rm d}t',
\end{equation}
where $\delt$ represents the last time-step.
The evaluation of the first integral can be done by using a simple trapezoidal rule and only the values of $\DMstarp$ at $t$ and $t-\delt$ are required.
The second integral is transformed by adding and subtracting $\delt$ in the exponential term, and simple algebra finally shows that
\begin{equation}
\label{dMwindfin}
  F_i(t)= \frac{\delt}{2}  \left[\DMstarp(t) + e^{-\frac{\delt}{b_i}}\DMstarp(t-\delt) \right] + e^{-\frac{\delt}{b_i}}F_i(t-\delt).
\end{equation}
Therefore, the introduction of the multi-exponential fit allow us to compute the instantaneous mass return rate at time $t$ just by storing the values of the $F_i$ functions at the previous time-step.
We performed a non-linear fit of the tabulated values of Eq.~(\ref{Wdef}) (obtained with the exact formulae).
We found that an acceptable approximation over the entire Hubble time is obtained with the coefficients in the first row of Table \ref{table:fit} (with a percentage error of at most of 10\% and only at very late times, when the mass return rate of a stellar population is negligible).

\subsection{SN Ia}
\label{appendixA2}

In a given simple stellar population, the total (i.e. volume-integrated) energy release by SNIa is
\begin{equation}
\label{LSN}
 \LSN(t)=E_{\rm SN} \RSN(t),
\end{equation}
where $E_{\rm SN}=10^{51}$ erg is the fiducial energy released by SNIa event and $\RSN$ is the SNIa rate. From Eqs.~(11)-(12) in \citet{co07}, the SNIa rate for a stellar population of mass $M_\ast$ can be written as 
\begin{equation}
\label{RSN}
 \RSN(t)=\frac{1.6 \times 10^{-13}}{\NBsun}\frac{M_*}{\Msun} \left(\frac{t}{t_{\rm H}}\right)^{-s} \;\;\;\qquad[\yrs^{-1}],
\end{equation}
where we adopt a stellar mass-to-light ratio $\NBsun=5$, $\theta_{\rm SN}=1$, $h=0.7$, $t_{\rm H}=13.7$ Gyrs, and $s=1.1$.
In our model with a star formation rate of $\DMstar$, it follows that in each time-step an amount $\DMstar\delt$ of stars is added to the galaxy body. Therefore, the instantaneous total rate of SNIa explosion at time $t$ is
\begin{equation}
\label{RSNT}
 \RSN^{\rm T}(t)=\frac{1.6 \times 10^{-13}}{\NBsun \Msun} \int_0^t \DMstar(t') \left(\frac{t-t'}{t_{\rm H}}\right)^{-1.1}{\rm d}t'.
\end{equation}
We expanded the dimensionless power-law kernel in the integral above again using a (dimensionless) multi-exponential function, so that all the considerations in Sect. \ref{appendixA1} hold and in particular Eq.~(\ref{dMwindfin}). The functions $F_i$ now contain new values of $b_i$ (in units of Gyrs$^{-1}$).
The parameters of the multi-exponential fit (with percentual errors $<3\%$ over 13.7 Gyrs) are listed in the second row of Table \ref{table:fit}.
Therefore, 
\begin{equation}
 \LSN(t)=  \frac{1.6\times 10^{38}}{\NBsun\Msun} \sum_i \frac{F_i}{a_i} \quad\quad\quad   \rm{erg\; s^{-1}}
\end{equation}
and finally, the average SNIa heating per unit time and unit volume, needed in the gas energy equation is
\begin{equation}
 \dot E_{\rm H,SNIa}=\frac{3\LSN(t)}{8\pi \rg^3},
\end{equation}
where again we use half-mass values.

\subsection{Column density for the Jaffe model}
\label{appendixA3}

In our model the gas density at each computation time is assumed to be proportional to $\rhoh$ and given by eq (\ref{rhogas}).
The corresponding projected gas distribution is
\begin{equation}
\label{sigmastar1}
     \begin{array}{l}
  \displaystyle \Sigmastar  \equiv 2\int_{R}^{\infty} \frac{\rho_{\rm g}(r)r \dr}{\sqrt{r^2-R^2}}. 
      \end{array}
\end{equation}
We note that the projected density formally diverges for $R\rightarrow0$, so that, to estimate a realistic value of the column density we need to compute a mean value of the projected density within a fiducial aperture.
The explicit formulae for $\Sigmastar$, and the associated projected mass encircled with $R$, $\MPg$, are given in Jaffe (1983). In particular, to estimate $\NH$ in Eq.~(\ref{NHdef}) we consider $\MPg$ 
\begin{equation}
\label{Mproj}
     \begin{array}{l}
 \displaystyle \MPg =2\pi\int_{0}^{R} R' \,\Sigma_{{\rm g}}(R'){\rm d}R' =\\
 \displaystyle \qquad \; \;\;\; =\Mg-4\pi\int_{R}^{\infty} r \rho_{\rm g}(r) \sqrt{r^2-R^2} {\rm d}r.
      \end{array}
\end{equation}
Integration of equation (\ref{Mproj}) yields
\begin{equation}
\label{Mprojsol}
\begin{array}{l}
\end{array}
\frac{\MPg}{M_{\rm g}}=\Rtilde \left\{ 
\begin{array}{l}
 \displaystyle \frac{\pi}{2}-\frac{\Rtilde ~{\rm arcsech}\Rtilde}{\sqrt{1-\Rtilde^2}}, 
         \; 0<\Rtilde<1; \\
 \\
 \displaystyle \frac{\pi}{2}-\frac{\Rtilde ~{\rm arcsec}\Rtilde}{\sqrt{\Rtilde^2-1}},                \; \Rtilde>1; \\
\end{array}\right.
\end{equation}
where $\Rtilde\equiv R/\rg$ and $M_{\rm P {\rm g}}(r_{\rm g})=(\pi/2-1)\Mg$.


\begin{thebibliography}{50}

\bibitem[\protect\citeauthoryear{Alexander et al.}{2003}]{a+03} Alexander D.~M. et al., 2003, AJ, 125, 383 

\bibitem[Alexander(2009)]{a+09} Alexander D.~M.,\ 2009, arXiv:0911.3911  

\bibitem[Ballero et al.(2008)]{2008A&A...478..335B} Ballero S.~K., Matteucci F., Ciotti L., Calura F., \& Padovani P.,\ 2008, \aap, 478, 335 

\bibitem[Bondi(1952)]{1952MNRAS.112..195B} Bondi H.,\ 1952, \mnras, 112, 195

\bibitem[Bregman 
\& Parriott(2009)]{2009ApJ...699..923B} Bregman J.~N., \& Parriott J.~R.,\ 2009, \apj, 699, 923 

\bibitem[Ciotti et al.(1991)]{1991ApJ...376..380C} Ciotti L., D'Ercole A., Pellegrini S., \& Renzini A.,\ 1991, \apj, 376, 380 

\bibitem[\protect\citeauthoryear{Ciotti \& Ostriker}{1997}]{co97} Ciotti L., Ostriker J.~P., 1997, ApJ, 487, L105

\bibitem[\protect\citeauthoryear{Ciotti \& Ostriker}{2001}]{co01} Ciotti L., Ostriker J.~P., 2001, ApJ, 551, 131

\bibitem[\protect\citeauthoryear{Ciotti \& Ostriker}{2007}]{co07} Ciotti L., Ostriker J.~P., 2007, ApJ, 1038, 1056

\bibitem[Ciotti et al.(2009)]{2009ApJ...699...89C} Ciotti L., Ostriker J.~P., \& Proga D.,\ 2009, \apj, 699, 89 
\bibitem[Ciotti et al.(2010)]{2010ApJ...717..708C} Ciotti L., Ostriker J.~P., \& Proga D.,\ 2010, \apj, 717, 708 

\bibitem[Cisternas et al.(2010)]{2010arXiv1009.3265C} Cisternas, M., et al.\ 2010, arXiv:1009.3265 

\bibitem[Comastri(2004)]{2004ASSL..308..245C} Comastri A.,\ 2004, Supermassive Black Holes in the Distant Universe, 308, 245 

\bibitem[Donley et al.(2010)]{2010ApJ...719.1393D} Donley J.~L., Rieke G.~H., Alexander D.~M., Egami E., \& P{\'e}rez-Gonz{\'a}lez P.~G.,\ 2010,\apj, 719, 1393 


\bibitem[\protect\citeauthoryear{Ferrarese \& Merritt}{2000}]{fm00} Ferrarese L., Merritt D., 2000, ApJ, 539, L9

\bibitem[\protect\citeauthoryear{Fukue}{2001}]{fukue01} Fukue J., 2001, PASJ, 687, 692

\bibitem[\protect\citeauthoryear{Haiman, Ciotti \& Ostriker}{Haiman et al.}{2004}]{hco04} Haiman Z., Ciotti L., Ostriker J.P., 2004, ApJ, 606, 763  

\bibitem[\protect\citeauthoryear{Hamann \& Ferland}{1999}]{hf99} Hamann F., Ferland G., 1999, ARA\&A, 37, 487

\bibitem[\protect\citeauthoryear{Heckman et al.}{2004}]{hkb+04} Heckman T.M., Kauffmann G., Brinchmann, J., Charlot S., Tremonti C., White S.D.M., 2004, ApJ, 613, 109

\bibitem[Hopkins et al.(2005)]{H2005} Hopkins, P.F., Hernquist, L., Cox, T.J., Di Matteo, T., Martini, P., Robertson, B., \& Springel, V.\ 2005, \apj, 630, 705 

\bibitem[Hopkins et al.(2006)]{H2006} Hopkins, P.F., Hernquist, L., Cox, T.J., Di Matteo, T., Robertson, B., \& Springel, V.\ 2006, \apjs, 163, 1 

\bibitem[Jaffe(1983)]{1983MNRAS.202..995J} Jaffe, W.\ 1983, \mnras, 202, 995

\bibitem[Johansson et al.(2009)]{2009ApJ...697L..38J} Johansson, P.H., Naab, T., \& Ostriker, J.~P.\ 2009, \apjl, 697, L38 

\bibitem[Kaviraj et al.(2010)]{2010arXiv1008.1583K} Kaviraj, S., Schawinski, K., Silk, J., \& Shabala, S.~S.\ 2010, arXiv:1008.1583

\bibitem[Kelly et al.(2010)]{2010ApJ...719.1315K} Kelly, B.~C., Vestergaard, M., Fan, X., Hopkins, P., Hernquist, L., \& Siemiginowska, A.\ 2010, \apj, 719, 1315

\bibitem[Krolik(1999)]{1999agnc.book.....K} Krolik, J.H.\ 1999, Active galactic nuclei: from the central black hole to the galactic environment, Princeton University Press

\bibitem[\protect\citeauthoryear{Kroupa}{2005}]{kr05} Kroupa P., 2001, MNRAS, 231, 246

\bibitem[\protect\citeauthoryear{Lanzoni et al.}{2004}]{lanz2004} Lanzoni B., Ciotti L., Cappi A., Tormen G., Zamorani G., 2004, ApJ, 640, 649

\bibitem[\protect\citeauthoryear{Lehmer et al.}{2004}]{l+04} Lehmer B.D. et al., 2004, astro-ph/0409600

\bibitem[\protect\citeauthoryear{Magorrian et al.}{1998}]{mtr+98} Magorrian J., Tremaine S., Richstone D., Bender R., Bower G., Dressler A. et al., 1998, AJ, 115, 2285 

\bibitem[\protect\citeauthoryear{Maraston}{2005}]{mar05} Maraston C., 2005, MNRAS, 799, 825

\bibitem[Marconi \& Hunt(2003)]{2003ApJ...589L..21M} Marconi, A., \& Hunt, L.~K.\ 2003, \apjl, 589, L21 

\bibitem[\protect\citeauthoryear{Mathews \& Bregman}{1978}]{mb78} Mathews W.~G., Bregman J.~N., 1978, ApJ, 224, 308

\bibitem[Mathews(1983)]{1983ApJ...272..390M} Mathews W.~G.,\ 1983, \apj, 272, 390 

\bibitem[Matteucci(2008)]{2008arXiv0804.1492M} Matteucci, F.\ 2008, arXiv:0804.1492 

\bibitem[Risaliti et al.(1999)]{1999ApJ...522..157R} Risaliti, G., Maiolino, R., \& Salvati, M.\ 1999, \apj, 522, 157 


\bibitem[\protect\citeauthoryear{Sazonov et al.}{2005}]{sos05} Sazonov S.Yu., Ostriker J.P., Ciotti L., Sunyaev R.A., 2005, MNRAS, 168 (SOCS)

\bibitem[Schawinski et al.(2009)]{2009ApJ...690.1672S} Schawinski, K., et 
al.\ 2009, \apj, 690, 1672 

\bibitem[\protect\citeauthoryear{Steidel et al.}{2002}]{s+02} Steidel C.C. et al., 2002, ApJ, 576, 653

\bibitem[\protect\citeauthoryear{Taam et al.}{1991}]{taamfu91} Taam R.E. and Fu A., 1991, ApJ, 696, 707

\bibitem[Treister et al.(2010)]{2010arXiv1009.2501T} Treister, E., Urry, C.~M., Schawinski, K., Cardamone, C.~N., \& Sanders, D.\ 2010, arXiv:1009.2501

\bibitem[\protect\citeauthoryear{Tremaine et al.}{2002}]{tgb+02} Tremaine S., Gebhardt K., Bender R., Bower G., Dressler A., Faber S.M. et al., 2002, ApJ, 574, 740 

\bibitem[\protect\citeauthoryear{Yu \& Tremaine}{2002}]{yt02} Yu Q., Tremaine S., 2002, ApJ, 335, 965 

\bibitem[Wild et al.(2010)]{2010arXiv1002.3156W} Wild, V., Heckman, T., \& Charlot, S.\ 2010, arXiv:1002.3156 

\end{thebibliography}
\end{document}